\documentclass[aps,pre,showpacs,superscriptaddress,twocolumn,floatfix]{revtex4}
\usepackage{graphicx}
\newcommand{\rt}{\rightarrow}
\def\be{\begin{equation}}
\def\ee{\end{equation}}
\def\bea{\begin{eqnarray}}
\def\eea{\end{eqnarray}}

\begin{document}

\title{On the equivalence of the freely cooling granular gas to the 
sticky gas}

\author{Mahendra Shinde}

\email{shinde_m_l@iitb.ac.in}

\affiliation{Department of Physics, Indian Institute of Technology, 
Bombay, Powai, Mumbai-400 076, India}

\author{Dibyendu Das}

\email{dibyendu@phy.iitb.ac.in}

\affiliation{Department of Physics, Indian Institute of Technology, 
Bombay, Powai, Mumbai-400 076, India}

\author{R. Rajesh}

\email{rrajesh@imsc.res.in}

\affiliation{Institute of Mathematical Sciences, CIT campus, Taramani, 
Chennai-600113, India}

\date{\today}

\begin{abstract} 
A freely cooling granular gas with velocity dependent restitution
coefficient is studied in one dimension. The restitution coefficient
becomes near elastic when the relative velocity of the colliding
particles is less than a velocity scale $\delta$. Different
statistical quantities namely density distribution, occupied and empty
cluster length distributions, and spatial density and velocity
correlation functions, are obtained using event driven molecular
dynamic simulations. We compare these with the corresponding
quantities of the sticky gas (inelastic gas with zero coefficient of
restitution). We find that in the inhomogeneous cooling regime, for
times smaller than a crossover time $t_1$ where $t_1 \sim \delta^{-1}$, 
the behaviour of the granular gas
is equivalent to that of the sticky gas.  When $\delta \rt 0$, then
$t_1 \rt \infty$ and hence, the results support an earlier claim that
the freely cooling inelastic gas is described by the inviscid Burgers
equation. For a real granular gas with finite $\delta$, the existence
of the time scale $t_1$ shows that, for large times, the granular
gas is not described by the inviscid Burgers equation.
\end{abstract}


\maketitle

\section{Introduction}

Granular systems, ubiquitous in nature, exhibit a wide variety of 
physical phenomena \cite{jaeger1996,kadanoff1999}. The non-equilibrium 
nature of these systems, characterized by dissipative particle 
collisions and external driving, has attracted a lot of attention from 
the theoretical point of view \cite{aranson2006}.  However, analytical 
progress remains slow. It is therefore important to understand 
completely simple models that capture some essential feature of these 
systems.

A simple, well studied model that captures dissipation effects, is the
freely cooling gas wherein particles at high temperature come to rest
through inelastic collisions
\cite{frachebourg,bennaim1,bennaim2,cattuto,puri, lattice1a,lattice1b,
lattice2,carnevale,kida,puglisi1,puglisi2,trizac1,trizac2,trizac3,mcnamara,goldhirsch,petzschmann,noije,brey1}.
At short times, the particles undergo homogeneous cooling, described
by the Haff's law \cite{haff}, wherein the kinetic energy of the
system decreases with time $t$ as $t^{-2}$.  This behaviour has
recently been verified experimentally \cite{haffexperi}.  Beyond a
crossover time $t_c$, the system crosses over to the inhomogeneous
cooling state, characterized by clustering, where the energy decays
according to a different power law $\sim t^{-\psi}$.  The exponent $\psi$
is dependent on dimension. For instance, $\psi=2/3$ in one dimension
and $\psi \approx 1$ in two dimensions.

In one dimension, it has been possible to make considerable
analytic and numerical progress.  
In the special case when the coefficient of restitution 
(henceforth denoted by $r$) is zero, {\it i.e.} particles stick on 
collision, the problem is exactly solvable. The problem can be mapped 
on to the description of shocks in velocity field in the inviscid 
Burgers equation \cite{frachebourg}. 
Burgers equation being exactly solvable  
in one dimension \cite{burgersbook}, 
the exact 
expressions for different quantities like velocity distribution, mass 
distribution, energy decay as a function of time 
are known.
This limit ($r=0$) will be referred to as the sticky gas.
In this paper, we will refer to the case $0 < r <1$ as the granular gas. 

Extensive simulations of the granular gas in one dimension showed
that, beyond the crossover time $t_c$, the energy decay in the
granular gas is identical to that of the sticky gas \cite{bennaim1}.
Not only is the exponent $\psi$ the same, all the curves for
different $r$'s collapse on top of each other. In addition, the
velocity distribution was seen to be independent of $r$. Finally, the
coarsening length scale ${\cal L}(t)$ scaled with $t$ identically for
both the granular and the sticky gas. It was thus concluded that the
long time behaviour ($t>t_c$) of the granular gas is described by the
Burgers equation. This analogy was generalized to higher dimensions
in Ref.~\cite{bennaim2}, though its applicability
remains unclear \cite{trizac2}.

To make this equivalence in one dimension stronger,
the two point correlations of the granular and the sticky
gases need to be compared.
In a recent paper \cite{Mahendra} we showed that, when $r$
depends on the relative velocity, there is a new crossover time $t_1$
(with $t_1>t_c$), beyond which the coarsening properties of the
granular gas differ from that of the sticky gas or the Burgers
equation. This was shown numerically by measuring the spatial
density-density and velocity-velocity correlation
functions. In this paper, we focus on the intermediate time
$t_c<t<t_1$, and show that the correlation functions of the granular
gas compare well with that of the sticky gas, obeying what is known as
the Porod law. In Sec.~\ref{sec:porod}, we review this law in the
context of coarsening systems, and point out some non-equilibrium
systems where it is known to be violated.

Our study involves event driven molecular
dynamics simulations of the granular and the sticky gases on a
ring. The precise definition of the model and the definition of the
statistical quantities that we measure are presented in
Sec.~\ref{sec:model}. The quantities that will be studied are 
density distributions, density-density and velocity-velocity 
correlation functions, and size distribution of empty and occupied clusters. 

Modeling the coefficient of restitution $r$ is crucial in defining the model
and understanding cooling granular gases. 
Experimentally, it is known that $r$ 
tends to $1$ (elastic limit) when the relative velocity of 
collision decreases to zero \cite{Raman,others1,others2}, while for 
large relative velocities $r$ tends to a constant. 
This introduces a velocity scale $\delta$ in the problem,
which consequently gives rise to a new time 
scale $t_1$. As in Ref.~\cite{Mahendra}, we model 
$r$ as
\be 
r = (1 - r_0) {\rm exp}\left(-|v_{\rm 
rel}/\delta|^{\sigma}\right) + r_0,
\label{eq:r} 
\ee
where $\delta$ represents a velocity scale, below 
which collisions become near elastic, {\it i.e}, when the relative
velocity $v_{rel} \ll \delta$, then
$r \rt 1$ (elastic collision). For $v_{rel} \gg \delta$ the 
collisions are inelastic with $r \rt r_0$. The parameter 
$\sigma$ decides how abruptly this crossover occurs. For example, for 
$\sigma \rt \infty$, $r$ versus $v_{rel}$ is a step function while for 
finite $\sigma$ it is a smoother curve. For $\sigma < 1$, the curve has 
a cusp as it approaches $r=1$.  The experimental curves~\cite{Raman} 
suggest that the power $\sigma$ can vary over wide range. Within
the framework of viscoelastic theory, a power of $\sigma=1/5$ has 
been predicted~\cite{poschel,brilliantov2}. Apart from that, studies of 
velocity distribution in freely cooling granular gases with velocity 
dependent restitution coefficient have been done using techniques 
different from that in this paper, namely direct simulation monte carlo and 
viscoelastic theories~\cite{brey2,brilliantov1}. 
With $\sigma \rt \infty$ and $\delta$ held finite, our problem becomes 
same as the granular gas problem studied earlier in~\cite{bennaim1}. 

The existence of a new crossover time $t_1$ was briefly shown in our
earlier publication \cite{Mahendra}. In Sec.~\ref{sec:crossover}, we
present simulation results, showing explicitly the existence of $t_1$
for different $\delta$ and $\sigma$ and determine its dependence
on $\delta$ and $\sigma$. We then proceed to do a detailed comparison of the
early time behaviour of the granular gas and the sticky gas in 
Sec.~\ref{sec:comparison}. We do the comparison on the basis of
data for the various statistical quantities characterizing coarsening in
these systems. In Sec.~\ref{sec:iia}, the 
two point correlations are related to the empty and occupied cluster 
distributions through a mean field approximation. This derivation 
leads us analytically to the Porod law for these coarsening systems. 
Conclusions and discussions are presented in Sec.~\ref{sec:summary}.

\section{The Porod law \label{sec:porod}}

The freely cooling granular gas is an example of what is known as a 
coarsening or a phase ordering system. It is well known that in systems 
freely relaxing to ordering states, there exists a length scale ${\cal 
L}(t)$ that increases with time~\cite{bray} . In addition, for usual 
phase ordering systems, the presence of a dominant ${\cal L}(t)$ results 
in a robust scaling law called the Porod law~\cite{bray,porod}. For 
scalar order parameters, the Porod law states that in $d$-dimensions, 
the scaled structure function $S(k,t)/{\cal L}^d \sim (k{\cal 
L})^{-{\theta}}$ for large $k{\cal L}$, with $\theta=d+1$. Hence, in one 
dimension $\theta=2$. In the above, 
$S(k,t)$ is the Fourier transform of the two 
point correlation function of the order parameter. We shall refer to 
systems having the above features as clean phase ordering systems. 
The essential fact is that, in clean phase ordering systems the relative fluctuation with 
respect to the ordered state is nominal. For example, in the case of a 
ferromagnetic Ising model, a finite system at infinitely large time 
shows magnetization per site peaked at two possible values of 
spontaneous magnetization.

A few non-equilibrium systems, e.g., particles sliding on a randomly
fluctuating surface~\cite{Dibyendu,manoj,nagar}, active nematics
namely, agitated granular-rod monolayers or films of orientable
amoeboid cells \cite{Ramaswamy}, show a different kind of phase
ordering; the main characteristic of which is undamped relative
fluctuations even in the thermodynamic limit. Hence, it is called
fluctuation dominated phase ordering. In such systems strong
fluctuations sustain in time without losing macroscopic order. The
order parameters have broad probability distributions even in the
thermodynamic limit indicating strong variation of the order parameter
in time. Unlike usual phase ordering, there need not be sharp
interfaces distinguishing one phase from the other. The
ensemble-averaged spatial correlation function shows a scaling form in
$|r/{\cal L}(t)|$, but unlike clean phase ordering systems, it exhibits either a cusp or
a power law divergence at small values of $|r/{\cal L}(t)|$. Hence the
scaled structure factor varies as $S(k,t)/{\cal L}^d \sim (k{\cal
L})^{-{\theta}}$ with $\theta<2$. Thus, measuring $\theta$ is a reliable
test to identify deviations from the Porod law.

We recently showed that the freely cooling granular gas at large
times $(t > t_1)$ also shows fluctuation dominated phase ordering \cite{Mahendra}.
The spatial correlation function shows a power law divergence for small values of $|r/{\cal
L}(t)|$. This in turn indicates violation of the Porod law in the
density and velocity structure functions with $\theta \approx 0.8$
\cite{Mahendra}. The Porod law violation here indicates the presence
of power law distributed clusters.  In other words the phase ordering
is {\it{not}} clean. Earlier, in a slightly different model in one dimension
 with velocity dependent restitution coefficient~ \cite{thiesen}, it was found that 
 the coarsening clusters are not long lived and they start breaking up at late times.
That result seems to have similarity to what we found in~\cite{Mahendra} at late times 
 $t > t_1$. But interestingly we find that at late times $(t > t_1)$ 
 there are two sub-regimes namely $t_2 > t > t_1$ and $t > t_2$. Here $t_2 \sim \delta^{-3}$
~\cite{bennaim1} is the very large scale beyond which all collisions become elastic and the system again 
 becomes homogeneous. On the other hand, the regime $t_2 > t > t_1$ is the new and 
 nontrivial regime of ``unclean" ordering that we have found~\cite{Mahendra} and are
 highlighting here.

 For Burgers equation, it was shown numerically in
Ref.~\cite{kida} [also see Sec.~\ref{sec:comparison}], that the ordering
is clean phase ordering, such that Porod law is obeyed. In this paper we examine the
structure factor for the granular gas for intermediate times and show
that Porod law is obeyed, like in the sticky gas.

\section{The model \label{sec:model}}

In this section, we define the model and the different physical quantities
that are  measured. Consider $N$ point particles of equal mass on a 
ring of length $L$. Initially, the particles are distributed randomly
in space with their velocities drawn from a normal distribution. Choose the
reference frame in which the center of mass momentum is zero. The 
particles undergo inelastic, momentum conserving collisions such that 
when two particles $i$ and $j$  with velocities $u_i$ and $u_j$ collide, 
the final 
velocities $u_i^{\prime}$ and $u_j^{\prime}$ are given by:
\be 
u_{i,j}^{\prime} =
u_{i,j} \left(\frac{1-r}{2} \right) + u_{j,i} \left(\frac{1+r}{2}\right),
\label{vupdate}
\ee 
where the relative velocity dependent restitution coefficient 
$r$ is given by Eq.~(\ref{eq:r}). For the sticky gas, $\delta=0$ and $r_0=0$,
such that $r=0$ for all collision velocities, and the particles coalesce on
collision.

For both the systems, the granular gas ($0<r_0<1$) and the sticky gas
($r=0$), we have done event driven molecular dynamics simulations with
system size $L=20000$ in units of the mean inter particle spacing at time
$t=0$.
The particle density is set to $1$ throughout.  Periodic
boundary conditions were imposed. For the granular gas, simulations
were performed for $r_0=0.1$, $0.5$ and $0.8$ in the time regime in
which ${\cal L}(t) \ll L$.  Thus, formally the sequence of limits are 
$L \rightarrow \infty$ first and then $t \rightarrow \infty$. In such a 
regime, the system continually coarsens, 
and hence we refer to it as a ``phase ordering system". 
In this time regime, simulations were done
for various different values of $\sigma$ (namely $3,4,5,10$ and
$\infty$), $\delta$ (namely $0.001$, $0.002$ and $0.004$). We found no
qualitative variation of our main results with varying values of $r_0,
\sigma$ and $\delta$.  Therefore, unless otherwise specified, the data
that is presented for the granular gas corresponds to $r_0 = 0.5,
\sigma=3.0$, and $\delta=0.001$.

It is well known that for $\delta = 0$, event driven molecular dynamics
simulations cannot go beyond the inelastic collapse. A similar difficulty is
faced when $\delta$ is finite, but $\sigma \rightarrow 0$.
For example, doing simulation with
realistic $\sigma$ like $1/5$ (suggested from viscoelastic theories)  or
$1/4$ ( for plastic collisions at large velocities), is out of our reach.
At such small $\sigma$ the shape of the $r$ versus $v_{rel}$ curve has a
cusp as $v_{rel} \rightarrow 0$. The particles practically do not see an
elastic regime and the simulation run into difficulties like particle overlap
and inelastic collapse.                                        
Nevertheless, from our studies of $\langle\rho^2\rangle$ versus $t$                      
(fig.3), we see that with decreasing $\sigma$ the ``clean phase ordering
regime" gets extended further in time. 
So we hope this trend would 
continue even for smaller $\sigma$ that our simulation could not access.
Also, the exponents are independent of the $\sigma$ values that we have
studied, and we hope this trend will hold for smaller $\sigma$ too.

The primary difference between a sticky gas and an granular gas is that 
in the former particles coalesce to form larger and larger mass 
clusters, while in the latter masses never coalesce. As a result, 
``clustering" phenomena in granular gas can be meaningfully quantified 
only if we define a local coarse-grained mass density $\rho$. To do so, 
first divide the ring into equally sized $N$ boxes. For the granular 
gas, $\rho_i$ is defined to be the total number of particles in the $i$th 
($i = 1,2,3,\ldots,N$) box. The coarse grained velocity $v_i$ 
is defined as the mean of the velocities of all the particles in box $i$.

We now define the statistical quantities that are measured for the
granular gas.  The extent of clustering can be measured in terms of
the variation of the density distribution $P(\rho,t)$. It is the
probability of finding a spatial box with density $\rho$ at time
$t$. The first moment $\langle \rho \rangle(t)$ of the distribution
$P(\rho,t)$ is constant at all times as the total mass is a conserved
quantity. The second moment $\langle\rho^2\rangle(t)$, is defined in
our simulation as ${\langle \sum_i {\rho_i}^2 \rangle}/N$, where the
sum is over all the boxes at time $t$, and the average
$\langle~\rangle$ is over initial conditions.

Other measures of clustering are the cluster lengths and gap
lengths. 
The distribution $O(x,t)$ is the number of clusters with exactly $x$
successive occupied boxes at time $t$, normalized by the total number 
of occupied clusters. Likewise, the
distribution $E(x,t)$ is the number of clusters with exactly $x$
successive empty boxes at time $t$, normalized by the total number of 
empty clusters. 

The other quantities of interest in this paper are spatial
correlation functions. The equal time, two-point density-density
spatial correlation function is defined as $C_{\rho\rho}(x,t) =
\langle \rho_i(t) \rho_{i+x}(t) \rangle$, where $\rho_i$ and
$\rho_{i+x}$ are densities at box $i$ and box $(i+x)$
respectively. The structure function $S_{\rho\rho}(k,t)$ is the
Fourier transform of $C_{\rho\rho}(x,t)$. Similarly, the equal time,
two-point velocity-velocity spatial correlation function is defined as
$C_{vv}(x,t) = \langle v_i(t) v_{i+x}(t) \rangle$, where $v_i$ and
$v_{i+x}$ are the coarse grained  velocities of  box $i$ and box
$(i+x)$ respectively. The structure function $S_{vv}(k,t)$ is the
Fourier transform of $C_{vv}(x,t)$.

In comparison to the granular gas, as particles coalesce in a sticky
gas, they form bigger and bigger mass clusters localized 
at points in space, such that their total number steadily decreases
with time. 
Therefore, without coarse graining,
one can measure mass and velocity 
distribution functions, and the corresponding two-point correlation
functions in space. 
But, to maintain uniformity with the granular gas, we will
continue to use coarse grained densities and velocities for the sticky gas.

\section{The new crossover time $t_1$ \label{sec:crossover}}

\begin{figure} 
\includegraphics[width=\columnwidth]{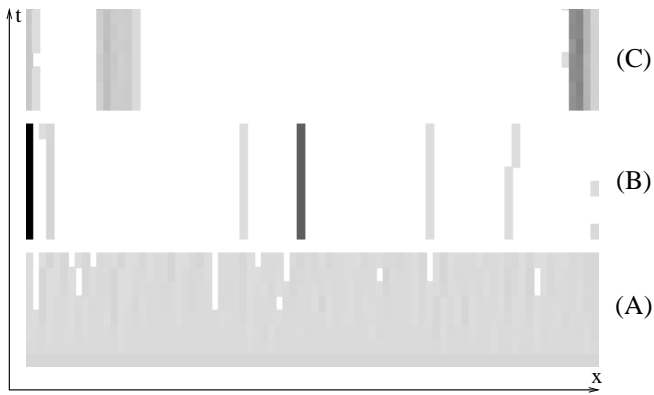} 
\caption{ 
\label{xt} The spatial density distribution is shown for different time
regimes: (A) $t = 0-7$ ($t \ll t_{c}$)
(B)$t = 600-607$ ($t_c < t < t_{1}$) and (C) $t = 8000-8007$ ($t
\gg t_{1}$). The darker regions correspond to higher densities. The data is
for the granular gas with $r=0.5$, $\delta=0.004$ and $\sigma=\infty$. For
these values of the parameters, $t_1 \simeq 1500$. 
}
\end{figure}

In this section, we show that the introduction of a velocity scale
$\delta$ introduces a new time scale $t_1$, beyond which the granular
gas deviates from the usual clean phase ordering as seen in the sticky
gas. We first show the different coarsening regimes by means of a space-time
density plot. In Fig.~\ref{xt}, the density distribution is shown for three
different time regimes with the darker regions corresponding to higher
density. 

In Fig~\ref{xt}(A), we show
a snapshot of the density profile of the system in a finite portion of
space for few successive time steps, for very early times,
$t \ll t_{c}$.  We see that the system is fairly homogeneous. In
Fig~\ref{xt}(B), a similar plot of the density profile is shown for
intermediate times $t_{c} < t < t_{1}$. The picture looks like any
ordinary clean phase ordering system with steady growth of isolated 
large density lumps (deep dark shaded), along with large empty gaps 
(white) separating them. If the clean phase ordering continues, then at later
times one would 
expect one or two very huge 
lumps remaining. Instead, in Fig.~\ref{xt}(C), we find grey
shaded spatially extended patches reappearing, along  with large white
gaps, implying fragmentation of some large clusters and
growth of smaller clusters.  This is the signature of fluctuation 
dominated phase ordering.

The above picture can be quantified by measuring the second moment of
the density distribution $\langle \rho^2 \rangle$. For the sticky gas,
it is known that $\langle \rho^2 \rangle \sim t^{2/3}$ for $t \gg 1$.
Does $\langle\rho^2\rangle$ in the granular gas behave as that of the
sticky gas? Since there are three parameters, $\sigma$, $\delta$ and
$r_0$ in the model, we need to carefully address the case with respect
to varying these parameters.
\begin{figure}
\includegraphics[width=\columnwidth]{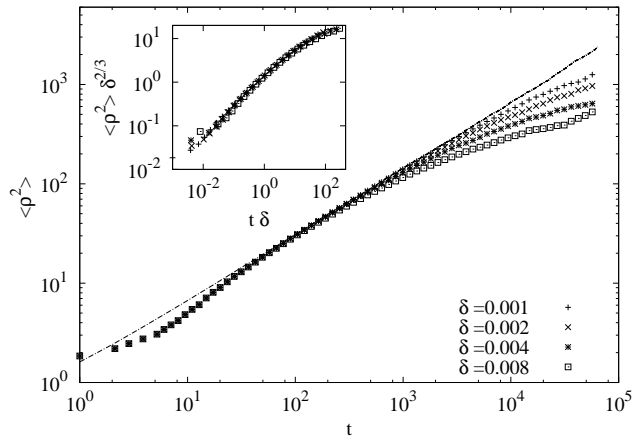}
\caption{ \label{S3deltas} The temporal variation of the second moment
of density, $\langle\rho^2\rangle$ for the granular gas, is shown for
different values of $\delta$ with $\sigma=3$ for all the cases. The data
is for the granular gas except for the dotted line which corresponds to
the sticky gas. The dotted line varies as $t^{2/3}$.
The data for the granular gas coincides with that of the sticky
gas for intermediate times, before deviating.
The inset shows data collapse
for various $\delta$ when the variables are scaled as in
Eq.~(\ref{sdeltas}).}
\end{figure}

Figure~\ref{S3deltas} shows the variation of $\langle\rho^2\rangle(t)$
with time $t$ for varying $\delta$, with $\sigma=3.0$ and $r_0=0.5$
fixed.  The trends are the same for other values of $\sigma$. For
early times $t_c <t < t_1$, $\langle\rho^2\rangle \sim t^{2/3}$ and
completely overlaps with the same data for the sticky gas (shown by a
line).  For late times $t \gg t_1$, $\langle\rho^2\rangle$ deviates
from the sticky gas behaviour, the crossover occurring later for
smaller $\delta$ (see Fig.~\ref{S3deltas}). The quantitative
dependence of the crossover time $t_1$ on $\delta$ can be obtained by
collapsing the different curves by scaling. The collapse is excellent
(see inset of Fig.~\ref{S3deltas}) when the different curves and $t$
are scaled as: \be \langle\rho^2\rangle = 
\frac{1}{\delta^{2/3}}f_{1}\left(t \delta \right).
\label{sdeltas}
\ee
Hence, we conclude that $t_1 \sim \delta^{-1}$.
		
\begin{figure} 
\includegraphics[width=\columnwidth]{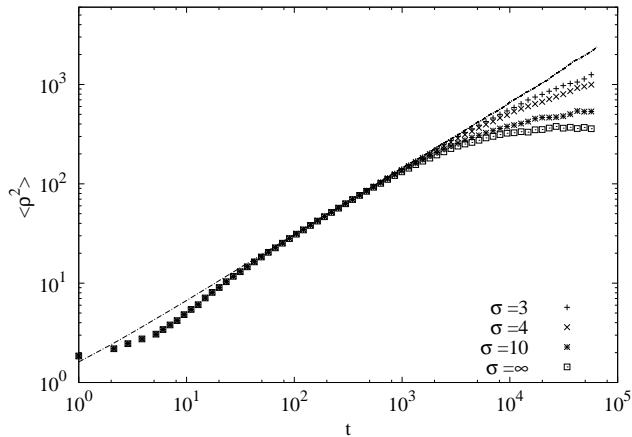} 
\caption{ \label{mntdel0.001} The temporal variation of the second
moment of density, $\langle\rho^2\rangle$ for the granular gas, is
shown for different values of $\sigma$ for $\delta=0.001$. The curve
with power $\langle\rho^2\rangle \sim t^{2/3}$ overlaps with all the
data for $t_c < t < t_1$. The crossover time $t_1$ increases with
decreasing $\sigma$, and is finite for $\sigma = \infty$.}
\end{figure}

The crossover time $t_1$ also depends on $\sigma$.
Figure~\ref{mntdel0.001} shows the variation of $\langle \rho^2
\rangle$ with time $t$ for different values of $\sigma$, keeping
$\delta$ fixed at $0.001$ and $r_0 = 0.5$. As earlier, for intermediate times
the behaviour of $\langle\rho^2\rangle$ mimics that of the sticky gas,
while for large times, there is deviation beyond a scale $t_1$. The
crossover time increases with decreasing $\sigma$. When
$\sigma=\infty$, $t_1$ is finite but nonzero.  Thus, we 
conclude that for all values of $\sigma$, the crossover time
$t_1$ is finite and nonzero.

Finally, we have checked that for fixed $\delta$ and $\sigma$, curves
of $\langle\rho^2\rangle$ do not vary with $r_0$, thus implying 
that $t_1$ does not depend on $r_0$.  

The coarsening behaviour for times $t \gg t_1$ was discussed extensively in
Ref.~\cite{Mahendra}. In this paper, we restrict ourselves to the regime
$t_c < t <t_1$, and compare the coarsening behaviour of the granular gas 
to that of the sticky gas. We bring out the similarities of the two 
through extensive simulations, the  results of which are given below.

\section{Comparison of the granular gas to the sticky gas
\label{sec:comparison}}

In this section, we present results for the granular gas and the
sticky gas for the intermediate times $t_c<t<t_1$. We show that
they
compare very well, and thus conclude strongly in favor of the
equivalence of the two in this time regime.  The following
quantities are studied: the density distribution $P(\rho,t)$, the
occupied cluster distribution $O(x,t)$, the empty interval
distribution $E(x,t)$, density--density correlations $C_{\rho
\rho}(r,t)$ (in real space) and $S_{\rho \rho}(k,t)$ (in Fourier
space) and the velocity--velocity correlations $C_{v v}(r,t)$.
For the sticky gas, the
scalings of these distribution functions are known exactly through the
exact solution in~\cite{frachebourg} and earlier studies
\cite{kida}. However, some of the results, for example the
velocity-velocity correlations, are complicated expressions whose
behaviour is not easily visualized. In addition, no plots are provided
for ready reference. For completeness and future reference, we present
numerical data for the sticky gas generated from event driven
simulations. The results are, of course, in agreement with
~\cite{frachebourg,kida}.

\subsection{Density Distribution function \label{sec:P_density}}

\begin{figure}
\includegraphics[width=\columnwidth]{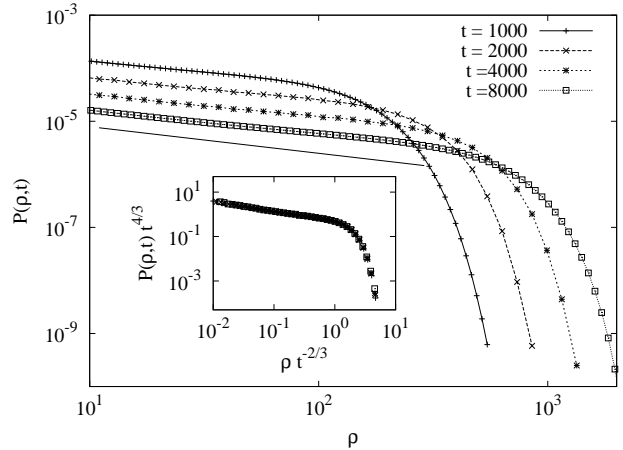}
\caption{ \label{0pmass2} The variation of $P(\rho,t)$ with $\rho$ is shown 
for different times for the sticky gas. The straight line has an exponent 
$-1/2$.  
The inset shows the data collapse when scaled as in Eq.~(\ref{0pms}).
}
\end{figure}

The mass or density distribution for the sticky gas is a power law with a
cutoff increasing with time, and an amplitude that decreases with time
\cite{frachebourg}. For the sticky gas, it has the scaling form
\be	
\lim_{L \rt \infty}P(\rho,t)= \frac{1}{t^{4/3}} f_{1}
\left(\frac{\rho}{t^{2/3}}\right),
\label{0pms}
\ee where the scaling function $f_1(z) \sim 1/\sqrt{z}$ when $z\ll 1$
and $f_1(z) \to 0$ when $z \gg 1$. Thus, for densities much smaller
than the cutoff, $P(\rho,t) \sim (t \sqrt{\rho})^{-1}$.  
The simulation results for the sticky gas are shown in
Fig.~\ref{0pmass2} for different times $t$. The data shows a time
dependent cutoff, and power law with exponent $-1/2$. The scaling
behaviour is consistent with Eq.~(\ref{0pms}) (see inset of
Fig.~\ref{0pmass2}).

\begin{figure} 
\includegraphics[width=\columnwidth]{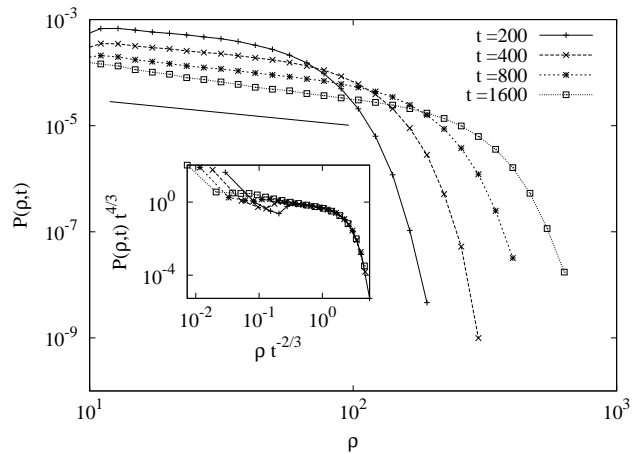} 
\caption{ 
\label{pd} The variation of the density distribution $P(\rho,t)$ with density
$\rho$ is shown for different times (in the early time regime) for the
granular gas.  The straight line has an exponent $-0.5$. The inset shows
the data collapse when scaled as in Eq.~(\ref{0pms}).}
\end{figure}

Now, consider the granular gas for $t_c < t < t_1$.  The variation of
the density distribution $P(\rho,t)$ with $\rho$ for different times
is shown in Fig.~\ref{pd}. The distribution is a power law with the
cutoff increasing with time and the amplitude decreasing with time.
The cutoff scales exactly as in the case of sticky gas, {\it i.e.},
$\rho_{\rm max} \sim t^{2/3}$.  A good data collapse (shown in the inset
of Fig.~\ref{pd}) is obtained when the different data are scaled
as in Eq.~(\ref{0pms}).

We note that if $\delta \rt 0$ the behavior of $P(\rho,t)$ as in
Fig.~\ref{pd} would continue indefinitely in time just as in sticky gas. 
But, in a more realistic granular gas model with finite $\delta$, for 
$t \gg t_1$, the behavior of $P(\rho,t)$ completely changes and 
a new scaling function emerges asymptotically as was detailed
in Ref.~\cite{Mahendra}. 

\subsection{Empty and Occupied cluster Distribution Functions 
\label{sec:EO_density}}

\begin{figure}
\includegraphics[width=\columnwidth]{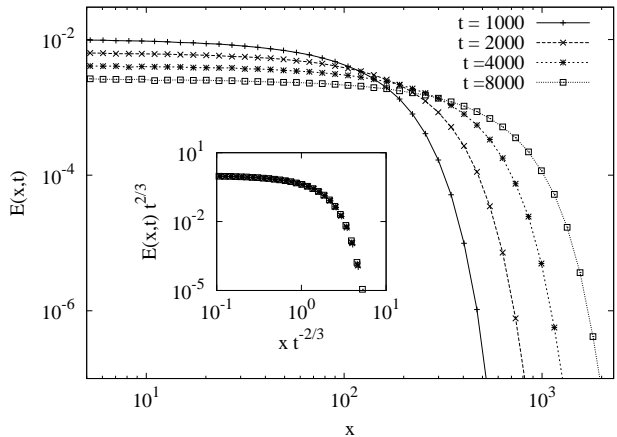}
\caption{ \label{0epgp} The variation of the empty interval distribution 
$E(x,t)$ with separation $x$ is shown for the sticky gas at different times. 
The inset shows the collapse of the curves when scaled as in
Eq.~(\ref{0epgn}).}
\end{figure}

A suitable measure of clustering in space is the 
empty gap distribution $E(x,t)$. For the sticky gas 
it is known \cite{frachebourg} that 
\be
E(x,t) = \frac{1}{t^{2/3}} f_{2} \left(\frac{x}{t^{2/3}}\right),
\label{0epgn}
\ee 
where the scaling function $f_2(z) \to {\rm constant}$ when  $z \to 0$ 
and $f_2(z) \to
0$ when $z \gg 1$. Thus, the largest empty cluster sizes $x_{\rm max} \sim
t^{2/3}$. The simulation data for $E(x,t)$ for the sticky gas is
plotted in Fig.~\ref{0epgp} for various times. The increasing cutoff
points to the steady increase in the size of empty clusters, as is
usual in phase ordering systems. The decreasing amplitude of the
distribution for small sizes indicate that small empty gaps are
steadily disappearing from the system with time. In the inset of
Fig.~\ref{0epgp}, the data collapse is shown following
Eq.~(\ref{0epgn}). 

\begin{figure} 
\includegraphics[width=\columnwidth]{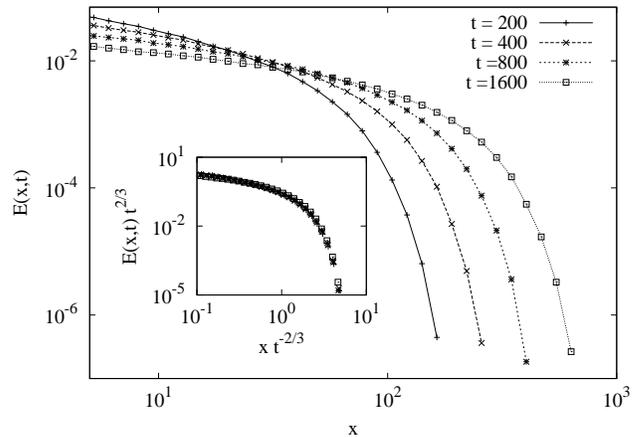} 
\caption{ \label{epgp} 
The variation of the empty interval distribution $E(x,t)$ with separation $x$
is shown for the granular gas 
at different times. 
The inset shows the collapse of the curves when scaled as in 
Eq.~(\ref{0epgn}).} 
\end{figure}

For the granular gas, the 
scaling and shape of the scaling function for $E(x,t)$ matches well
with that 
of the sticky gas. In Fig.~\ref{epgp} we present the data for various 
times and in the inset the scaling collapse using  Eq.~(\ref{0epgn}).
The collapse is excellent, showing the equivalence of the granular gas to the
sticky gas as far as $E(x,t)$ is concerned.

It turns out that in the granular gas for $t \gg t_1$, 
the above clean phase ordering 
behaviour of $E(x,t)$ dramatically changes and a power law
distribution for the scaling function $f_2(z)$ appears with a large
negative power ($\approx 2.2$) ~\cite{Mahendra}, signaling a
crossover to fluctuation dominated phase ordering behaviour.

\begin{figure}
\includegraphics[width=\columnwidth]{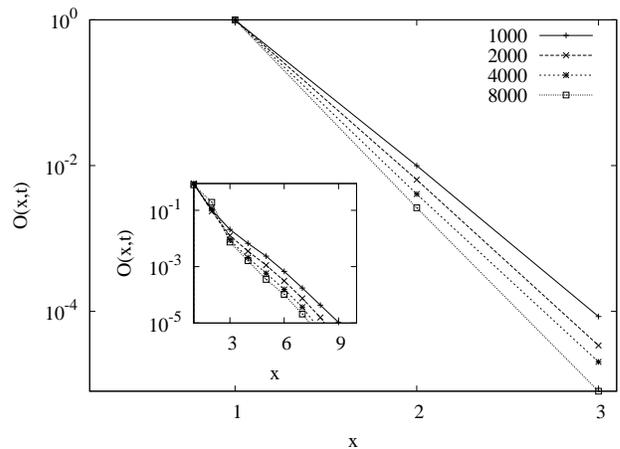}
\caption{ \label{0opgp} The variation of the occupied cluster length
distribution $O(x,t)$ with separation $x$, is shown at different times for
the sticky gas on a linear-log plot.
Inset: A similar plot for the granular gas. The times are 
$t = 200$(top curve), $400, 800$ and $1600$(bottom curve).
}
\end{figure}

The occupied cluster distribution $O(x,t)$, has a very different
form and evolution in comparison to $E(x,t)$ discussed above.  In the
absence of any analytical guidance on this quantity from earlier works
on sticky gas, we present directly what we find in our simulation. The data
for $O(x,t)$ is shown in 
Fig.~\ref{0opgp}. We see that large occupied clusters are extremely 
rare, and within size $3$ the distribution fall to very low value. 
The distribution function $O(x,t)$ is exponential with $x$, and 
its width decreases marginally with time. Thus, we conclude that 
occupied clusters are highly localized in space and 
do not vary much in time. 

In the inset of Fig.~\ref{0opgp}, we show $O(x,t)$ for the granular 
gas in the early time regime. The curves are strikingly similar to the 
sticky gas. There is an exponential decay over very short sizes,
and clusters beyond size $9$ are very rare. Again there is a marginal 
decrease in width of the distribution with time. Just as the sticky gas,
this implies extremely localized occupied clusters, whose width do
not vary appreciably with time. 

We note that, in sharp contrast to the above scenario, the 
occupied clusters do not remain localized at large times $t \gg t_1$,
for the granular gas. In fact $O(x,t)$ also decays as a power law
like $E(x,t)$ with the same negative power ($\approx 2.2$) as 
was shown in Ref.~\cite{Mahendra}.

\subsection{Density-density correlation functions \label{sec:C_rhorho}}

\begin{figure}
\includegraphics[width=\columnwidth]{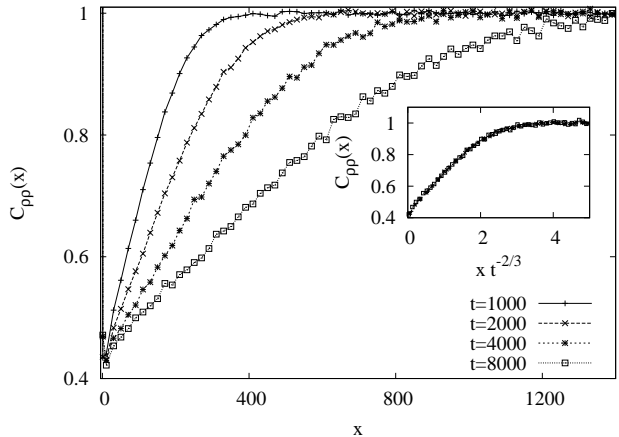}
\caption{ \label{0crho_inset} The density-density correlation function 
$C_{\rho\rho}$ as a function of $x$ for the sticky gas. 
Inset: Shows the data collapse when scaled as in
Eq.~(\ref{e0crho_inset}).}
\end{figure}

The two point spatial correlation functions give a very vivid picture
of the type of phase ordering that the system undergoes, and we
discuss these in this and the next subsection.  The data for two
point density-density correlation function $C_{\rho\rho}$ (defined in
Sec.~\ref{sec:model}) of the sticky gas is plotted in
Fig.~\ref{0crho_inset}. The correlation function  drops from a very large 
value at
$x=0$ to a value below $1$ and then asymptotically increases to $1$ as
$x \rt \infty$. The initial drop is because near every particle
there is a density depletion zone created by coalescence of clusters.
The asymptotic approach is expected as
$C_{\rho\rho} \rt {\langle \rho \rangle}^2$ for large $x$ and
${\langle \rho \rangle} = 1$. For a phase ordering system \cite{bray}
the scaling form
\be
C_{\rho\rho}(x,t) = f_3(x/{\cal L}),
\label{e0crho_inset}
\ee
is expected, and we show the data collapse in the inset of 
Fig.~\ref{0crho_inset}. For small
scaled distance $x/{\cal L}$, the scaling function $f_3$ increases   
linearly, implying Porod law \cite{porod} to be valid. 

\begin{figure}
\includegraphics[width=\columnwidth]{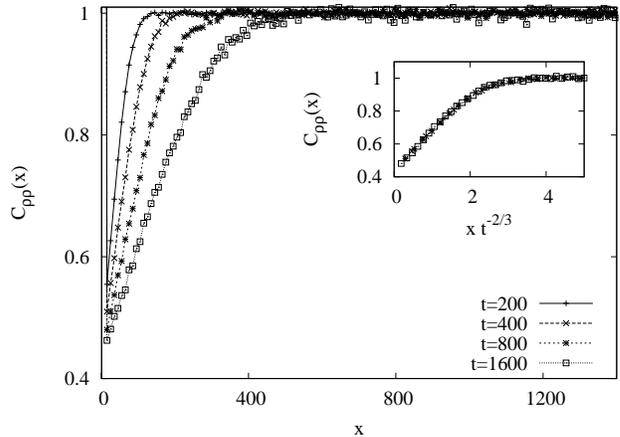}
\caption{ \label{crho_inset} The density-density correlation function
$C_{\rho\rho}$ as a function of $x$ for the granular gas. Inset:
Shows the data collapse when scaled as in Eq.~(\ref{e0crho_inset}).}
\end{figure}

The corresponding data for $C_{\rho\rho}$ for the granular gas at
intermediate times is shown in Fig.~\ref{crho_inset}. We find a
striking similarity between Fig.~\ref{crho_inset} and
Fig.~\ref{0crho_inset} (for the sticky gas).  The scaling function
shown in the inset of Fig.~\ref{crho_inset} is very similar to the
inset in Fig.~\ref{0crho_inset}.

\begin{figure}
\includegraphics[width=\columnwidth]{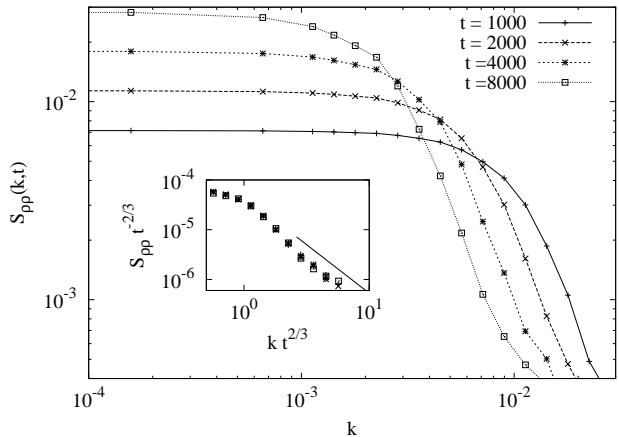}
\caption{ \label{0dcorr0G} The variation of ${S}_{\rho\rho}$ with $k$ 
for the sticky gas is shown for different times. The inset shows 
the data collapse when scaled as in  Eq.~(\ref{0dcorr}). The straight line 
has an exponent $-2.0$.}
\end{figure}

\begin{figure}
\includegraphics[width=\columnwidth]{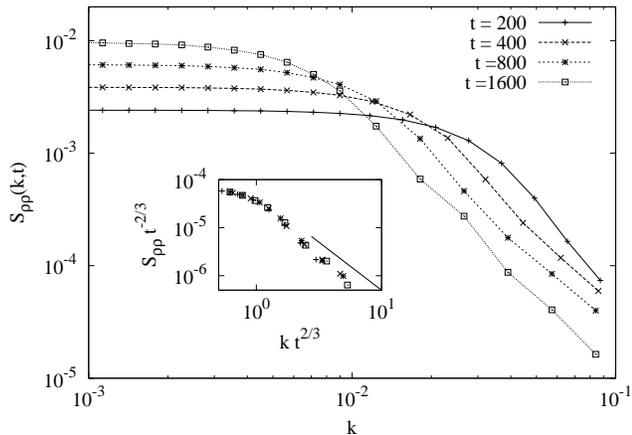}
\caption{ \label{dcorr} The variation of $S_{\rho\rho}$ with $k$ is 
shown for the granular gas at early times. The inset shows the data
collapse when scaled as in
Eq.~(\ref{0dcorr}). The straight line has an
exponent $-2.0$.}
\end{figure}

To find the structure function ${S}_{\rho\rho}(k,t)$ which
scales in the conventional fashion one has to treat the data for
$C_{\rho\rho}(x,t)$ carefully. Firstly, the value at $x/{\cal L} = 0$
for the scaled $C_{\rho\rho}(x,t)$ do not form a part of the scaling
function $f_3$ and have to be carefully adjusted by hand so as not to
introduce undesirable distortions in $k$-space. This procedure has
been discussed in details in \cite{Dibyendu}, and we do the same in
this case. Secondly, $f_3(z)$ does not decay to zero at $z \rt \infty$
and well behaved Fourier transform cannot be found. So for this
special case of the density-density correlation function for both the
sticky and the granular gases, we define ${S}_{\rho\rho}(k,t)$ to be
the Fourier transform of $1 - C_{\rho\rho}(x,t)$. The latter real
space function nicely decays to zero at large $x$ and thus
${S}_{\rho\rho}(k,t)$ is expected to be well behaved. The data thus
obtained are shown in Fig.\ref{0dcorr0G} for the sticky gas, and in
Fig.~\ref{dcorr} for the granular gas (for small times).  They
compare very well, and in the insets of Figs.~\ref{0dcorr0G} and
\ref{dcorr}, the scaling collapse of the data are shown according to
\be 
{S}_{\rho\rho}(k,t)={\cal L} g_{3}(k{\cal L}(t)).
\label{0dcorr}
\ee 
For large $k {\cal L}$ we see that for both the sticky and the
granular gas the scaling function $g_{3}(z) \sim z^{-2}$ for $z \gg
1$, i.e obey Porod law, as is expected for a clean phase ordering system.

For times $t \gg t_1$ there is a shift to non-Porod behaviour for the 
granular gas as discussed in Ref.~\cite{Mahendra}. The functional form 
of $g_3(x)$ completely changes in that regime.

\subsection{Velocity-velocity correlation functions \label{sec:C_vv}}

\begin{figure}
\includegraphics[width=\columnwidth]{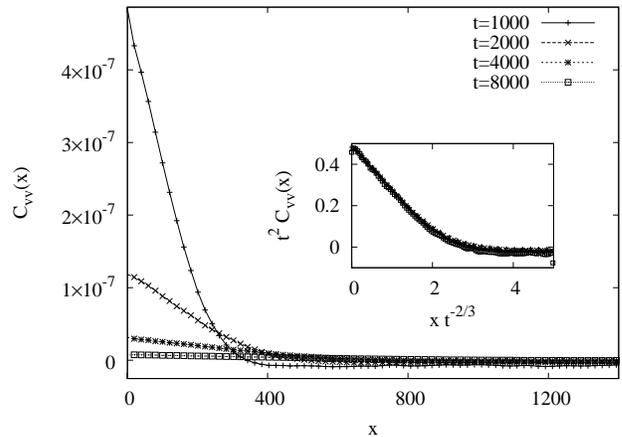}
\caption{ \label{0cvb_inset} The figure shows the variation of  $C_{vv}(x,t)$
with $x$ for different times. The data is for the sticky gas. 
Inset: shows the data collapse when scaled as in 
Eq.~(\ref{e0cvb_inset}).}
\end{figure}

The velocity-velocity correlation function $C_{vv}$ is expected
\cite{Mahendra} to scale as $\sim (v_t/{\cal L})^2 f_4(x/{\cal L})$, where
$v_t$ is the typical velocity which scales as $t^{-1/3}$, and the factor
$1/{\cal L}$ accounts for probability of a box being occupied. The
scaling function $f_4(x)$ 
decays to zero for large $x$. Putting in the $t$
dependence for $v_t$ and ${\cal L}$, we obtain
\be 
C_{vv}(x,t) = \frac{1}{t^{2}} f_{4}(x/{\cal L}).
\label{e0cvb_inset}
\ee
In Fig.~\ref{0cvb_inset}, the data for $C_{vv}(x,t)$ 
at different times is shown for the sticky gas. 
The inset shows the data collapse when scaled as in 
Eq.~(\ref{e0cvb_inset}). 
The scaling function $f_4(z)$ decays linearly for small $z$ 
and goes to zero for large $z$. The linearity is again a clear signature of 
the Porod law.

\begin{figure}
\includegraphics[width=\columnwidth]{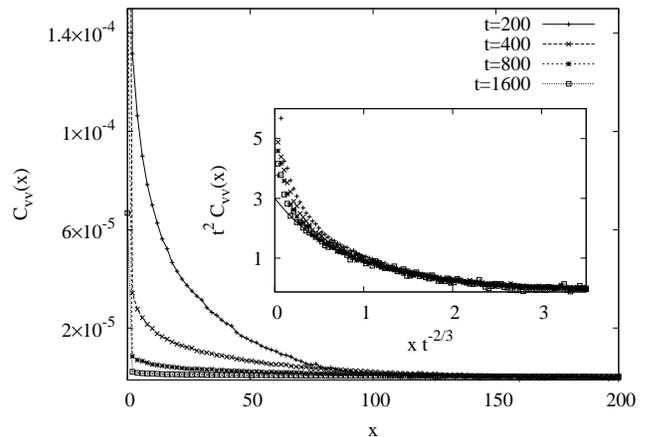}
\caption{ \label{cvb_inset} The figure shows the variation of 
$C_{vv}(x,t)$ with $x$ for different times for the granular gas. 
Inset: shows the curves when scaled as in
Eq.~(\ref{e0cvb_inset}). The higher curves correspond to smaller
times.  At large times, the data approaches the 
solid line, an exponential obtained by fitting the scaling region.
}
\end{figure}

The corresponding data for $C_{vv}(x,t)$ for the granular gas is shown
in Fig.~\ref{cvb_inset}. There are corrections to scaling
when compared to the corresponding data for the sticky gas (see inset of
Fig.~\ref{cvb_inset}). A possible
reason is that the times for
which the measurements are taken for the granular gas are a factor of
ten smaller that of the sticky gas. However, it is seen (see inset of
Fig.~\ref{cvb_inset}) that, for increasing time, the $x/{\cal L}$ at
which deviation from scaling occurs decreases. The scaling function
tends towards the curve $3.0 \exp(-1.25 x/{\cal L})$ (drawn
as a solid line in the inset of Fig.~\ref{cvb_inset})). Thus, again a
linear decay is obtained near the origin, signifying Porod law.
The scaling functions in the inset of Figs~\ref{0cvb_inset} and \ref{cvb_inset}
differ by a factor of $6$ for small $x/{\cal L}$. This is probably a
consequence of the coarsening rule that we have used.

We also note that
the structure function ${S}_{vv}(k,t)$, obtained as the Fourier 
transform of $C_{vv}(x,t)$, should scale as
\be
{S}_{vv}(k,t)=\frac{1}{t^{4/3}} g_4(k{\cal L}).
\label{0vcorr}
\ee
However, due to corrections to scaling for small $x/{\cal L}$, we cannot
obtain good data for large $k$, and hence we do not show the structure
functions as plots.

\section{The independent interval approximation \label{sec:iia}}

In this section, starting from the empty and occupied interval
distributions, we derive the Porod law for density-density
correlations via a mean field approximation, often referred to as
independent interval approximation \cite{satya}. This approximation
was used in Ref.~\cite{Mahendra} to connect the interval distributions
to the density correlations for times $t \gg t_1$.  Here, we first
give the derivation of the result used in Ref.~\cite{Mahendra}, and
then apply it to times $t_c<t<t_1$.
\begin{figure}
\includegraphics[width=\columnwidth]{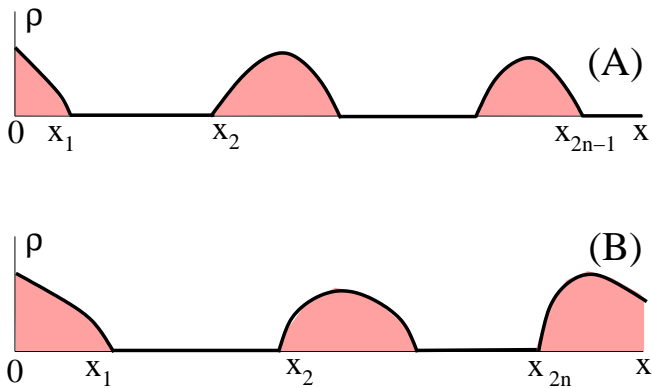}
\caption{ \label{fig:iia} A schematic diagram showing configurations that
contribute to $p_k$ when $k=2n-1$ is odd (A) and when $k=2n$ is even (B). The
shaded regions represent boxes with nonzero density $\rho$. The even
$k$ correspond to the case when both the sites $0$ and $x$ are occupied and
contribute to the density-density correlation function $C_{\rho\rho}(x,t)$.}
\end{figure}

The equal time density--density correlation function 
$C_{\rho\rho}(x) = \langle{\rho_i}(t)\rho_{i+x}(t)\rangle$
is the expectation value of the product of densities of boxes 
separated by spatial distance $x$. This quantity can be estimated using 
an approximation which is discussed below. Let $p_k(x)$ denote
the probability that given that site $0$ is occupied, there are
exactly $1+[k/2]$ occupied intervals and $[(k+1)/2]$ empty intervals, where
$[\ldots ]$ denotes the integer part. An example of configurations
contributing to $p_n$ is shown in  Fig.~\ref{fig:iia}(A) for odd $k=2n-1$ and
Fig.~\ref{fig:iia}(B) for even $k=2n$.
Thus, $p_0(x)$ will correspond to
the situation when $0$ is occupied and there is continuous set of
occupied sites up to $x$, while $p_1(x)$ would correspond to a set of
occupied sites starting from $0$, followed by an empty interval
extending beyond $x$. Since $C_{\rho\rho}(x)$ is approximately equal to
density square times the probability that there is a nonzero density
at $x$ given there is a nonzero density at $0$, we obtain
(setting $\rho=1$), 
\be
C_{\rho\rho}(x) \approx \sum_{n=0}^{\infty} p_{2n}(x), 
\label{eq:iia1}
\ee
where $p_{2n}(x)$ is the probability
of having exactly $n$ empty gaps between $0$ and $x$ with $0$ and $x$ being
occupied. 

The probabilities $p_n(x)$ may be expressed in terms of the 
gap distributions $E(x,t)$ and $O(x,t)$. Let 
\bea
Q(x,t) &=& \int_{x}^{\infty} dx' O(x',t), \label{eq:iia2}\\
F(x,t) &=& \int_{x}^{\infty} dx' E(x',t). \label{eq:iia3}
\eea
where $Q(x,t)$ is the probability of getting an occupied 
cluster of size greater than $x$. Similarly $F(x,t)$ is the probability 
of getting an empty cluster of size greater than $x$.

Approximating joint distributions $p_{2 n}(x)$ and $p_{2 n -1}(x)$ by products of
individual distributions of the intervals (namely $E$, $O$, $Q$ and $F$), one can easily 
verify (following Fig.~\ref{fig:iia} as guideline for the odd and the even cases), that
\begin{widetext}
\bea
p_{2 n -1}(x)& = & \alpha \int dx_1 \ldots dx_{2n-1}
Q(x_1) F(x-x_{2n-1})
\prod_{j=1,3,\ldots}^{2 n -3}
\left[E(x_{j+1}-x_j) O(x_{j+2}-x_{j+1})\right],
~~n=1,2,\ldots  \label{eq:iia4} \\
p_{2 n}(x)& = &\alpha  \int dx_1 \ldots dx_{2n}
Q(x_1) E(x_{2n}- x_{2n-1}) Q(x-x_{2n})
\prod_{j=1,3,\ldots}^{2 n -3}
\left[E(x_{j+1}-x_j) O(x_{j+2}-x_{j+1})\right],
~~n=1,2,\ldots  \label{eq:iia5} \\
p_{0}(x)& = &\alpha  \int dx' Q(x'),   \label{eq:iia6} 
\eea
\end{widetext}
where $\alpha$ is a normalization constant that will be determined
shortly.

Equations(\ref{eq:iia4})--(\ref{eq:iia6}) simplify considerably
if we take Laplace
transforms. Denoting the Laplace transform of a function $f(x)$ by
$\tilde{f}(s)$, where $\tilde{f}(s) = \int_0^{\infty} f(x) e^{-s x}
dx$, we obtain
\bea
\tilde{p}_{2 n -1}(s)& = & \alpha \tilde{Q} \tilde{F} (\tilde{E}
\tilde{O})^{n-1}, ~~n=1,2,\ldots  \label{eq:iia7} \\
\tilde{p}_{2 n}(s)& = &\alpha  
\tilde{Q}^2 \tilde{E} (\tilde{E}
\tilde{O})^{n-1}, ~~n=1,2,\ldots  \label{eq:iia8} \\
\tilde{p}_{0}(s)& = &\frac{\alpha}{s} \left [ \langle x\rangle_0 -
\tilde{Q} \right],   \label{eq:iia9} 
\eea
where $\langle x \rangle_0 = \int_0^\infty dx x O(x)$, is the mean length of
an occupied cluster. The distribution $\tilde{Q}(s)$ can
be expressed in terms of the probability density $\tilde{O}(s)$ as
\be
\tilde{Q} (s) = \frac{1}{s} \left[1-\tilde{O} (s) \right],
\label{eq:iia10}
\ee
and likewise for $\tilde{F}$ (with $\tilde{O}$ replaced in Eq.~\ref{eq:iia10} 
by $\tilde{E}$). The constant $\alpha$
can now be determined from the condition $\sum_{n=0}^\infty
\tilde{p}_n(s) = s^{-1}$, which follows from $\sum_{n=0}^\infty p_n(x) = 1$.  
We immediately obtain $\alpha=1/\langle x \rangle_o$. 
Equations~(\ref{eq:iia1}),(\ref{eq:iia7}),(\ref{eq:iia9}) and (\ref{eq:iia10}) 
give
\be
\tilde{C}_{\rho\rho}(s) = \frac{1}{s} - \frac{[1-\tilde{O}(s)]
[1-\tilde{E}(s)]}
{\langle x \rangle_o s^2 [1-\tilde{E}(s) \tilde{O}(s)]}.
\label{eq:iia11}
\ee

We now need the form of $\tilde{O}(s,t)$ and $\tilde{E}(s,t)$. The occupied
cluster size distribution $O(x,t)$ has no dependence on the length scale
${\cal L}(t)$ (see Fig.~\ref{0opgp}). 
The distribution is very close to a pure
exponential and we take it to be of the form $O(x,t) 
\approx a^{-1} \exp(-x/a) $
where $a$ is independent of time. Then $\tilde{O}(s,t) \approx
(a s+1)^{-1}$. The
empty interval distribution $E(x,t)$ is a constant for small $x$ and goes to
zero for $x \gg L_t$ (see Figs.~\ref{0epgp} and \ref{epgp}).
We approximate it by a step function: $E(x,t) = {\cal L}_t^{-1}$ for $x\leq
{\cal L}_t$, and $E(x,t) = 0$ for $x>{\cal L}_t$.
Then, $\tilde{E}(s,t) \approx (s{\cal L}_t)^{-1} [1-\exp(-s {\cal L}_t)]$.

We are interested in the limit $s\to 0$, ${\cal L} \to \infty$ keeping $s
{\cal L} $ a constant. For verifying Porod law, we are further interested in
the limit $s {\cal L} \gg 1$. Substituting the assumed forms for $\tilde{E}$
and $\tilde{O}$ in Eq.~(\ref{eq:iia11}) and expanding for large $s {\cal L}$,
we obtain
\be
\frac{C_{\rho\rho}(s)}{{\cal L}} \approx \frac{1}{(s {\cal L})^2}, \quad s
{\cal L} \gg 1.
\ee
We thus obtain the Porod law behaviour by using the cluster distributions.
The same independent interval approximation gives consistent results between
the cluster distributions and the two point correlation functions in late time
regime $t > t_1$ \cite{Mahendra}, where there is a violation of Porod law.

\section{Summary and conclusions \label{sec:summary}}

In this paper, we studied the density distribution, the empty and
occupied cluster distributions, the spatial density--density
correlations and the velocity--velocity correlations, in a granular
gas undergoing inelastic collisions in one dimension. The restitution
coefficient was modeled as being velocity dependent, with the
collisions becoming nearly elastic when the relative velocity is less
than a velocity scale $\delta$. The velocity scale $\delta$ introduced
a new time scale $t_1$ into the problem. In this paper, the dependence
of $t_1$ on the different parameters was found. Due to the existence
of $t_1$, and the crossover scale $t_c$ between the homogeneous and
inhomogeneous cooling regimes, there are three different time regimes
in the problem: (1) $t<t_c$, during which particles undergo
homogeneous cooling, (2) $t_c<t<t_1$, referred to as intermediate
times, and (3) $t>t_c$ referred to as late times. In this paper, the
focus was on the intermediate times.

For intermediate times, we compared the different correlation functions 
of the granular gas with that of the sticky gas. We found that there is an
excellent match between the two.
Coarsening in both the cases is
governed by the Porod law. These results thus support the claim that the
Burgers equation is the correct description for the granular gas
\cite{bennaim1}. 

For late times, this equivalence breaks down \cite{Mahendra}. The coarsening
in the granular gas violates Porod law. The occupied cluster distribution and
the empty cluster distributions differ significantly from intermediate times
by developing into power laws. The existence of clusters of all sizes leads
to fluctuation dominated coarsening. A continuum equation describing this
regime is missing, and it would be interesting to find one in future.

\begin{acknowledgments}
We thank M. Barma for useful discussions.  D.D. was supported by grant
no. $3404-2$ of ``Indo-French Center (IFCPAR)/ (CEFIPRA)''.
\end{acknowledgments}

%\bibliography{ref}

\end{document}